\documentclass[epj]{svjour}
\usepackage{graphics}
\usepackage{epstopdf}
\begin{document}
\title{Physical characterization of NEA Large Super-Fast Rotator (436724) 2011 UW158}
\author{A. Carbognani\inst{1} 
\and B. L. Gary \inst{2}
\and J. Oey \inst{3}
\and G. Baj \inst{4}
\and P. Bacci \inst{5}
}                     
\offprints{A. Carbognani}          
\institute{Astronomical Observatory of the Autonomous Region of Aosta Valley (OAVdA), Aosta - Italy 
\and Hereford Arizona Observatory (Hereford, Cochise - U.S.A.)
\and Blue Mountains Observatory (Leura, Sydney - Australia)
\and Astronomical Station of Monteviasco (Monteviasco, Varese - Italy)
\and Astronomical Observatory of San Marcello Pistoiese (San Marcello Pistoiese, Pistoia - Italy)}
\date{Received: date / Revised version: date}
%
\abstract{
Asteroids of size larger than 0.15 km generally do not have periods smaller than 2.2 hours, a limit known as cohesionless spin-barrier. This barrier can be explained by the cohesionless rubble-pile structure model. There are few exceptions to this ``rule'', called LSFRs (Large Super-Fast Rotators), as (455213) 2001 OE84, (335433) 2005 UW163 and 2011 XA3. The near-Earth asteroid (436724) 2011 UW158 was followed by an international team of optical and radar observers in 2015 during the flyby with Earth. It was discovered that this NEA is a new candidate LSFR. With the collected lightcurves from optical observations we are able to obtain the amplitude-phase relationship, sideral rotation period ($PS = 0.610752 \pm 0.000001$ h), a unique spin axis solution with ecliptic coordinates $\lambda = 290^{\circ} \pm 3^{\circ}$, $\beta = –39^{\circ} \pm 2^{\circ}$ and the asteroid 3D model. This model is in qualitative agreement with the results from radar observations.
\PACS{
      {PACS-key}{discribing text of that key}   \and
      {PACS-key}{discribing text of that key}
     } 
} 
\maketitle
\section{Introduction}
\label{intro}
The near-Earth asteroid (436724) 2011 UW158 was discovered on 2011 Oct 25 by the Pan-STARRS 1 Observatory at Haleakala (Hawaii, USA). This asteroid is classified as Apollo-type and, considering that the MOID (Minimum Orbit Intersection Distance), with Earth is 0.0026 AU it is also a PHA (Potentially Hazardous Asteroid)\cite{ipatov}. \\
A flyby with Earth occurred on 2015 July 19 at 0.0164 AU (Astronomical Unit), so 2011 UW158 also became a radar target for Goldstone, Arecibo and Green Bank \cite{naidu}. This asteroid was the subject of an international radar system collaboration because it has been illuminated with the Goldstone 70-m antenna and the signal reflected was received with the 32-m radio telescopes of the Quasar VLBI network at the Zelenchukskaya and Badary Observatories \cite{ipatov}. The radar results suggest that the body has a prolate shape with dimensions of about $600\times 300$ m, a spin period of $36 \pm 3$ min and an inhomogeneous surface \cite{naidu} \cite{ipatov}. This asteroid was also followed by an international team of optical observers (see Table \ref{ast}), for 31 nights between 2015 June 17 and Sept 26 (see Table \ref{amp} for the list of the observing sessions).\\ 
After a concise review about cohesionless spin-barrier we will see the discovery circumstance of 2011 UW158 fast rotation period, the amplitude-phase relationship and the pole search procedure. Finally, we will compare the shape from optical observations with the results from radar observations.  

\begin{table*}
\caption{Observers, telescopes and (unfiltered) CCD camera used for 2011 UW158.}
\begin{center}
\begin{tabular}{lll}
\hline
\\
Observer & Telescope & CCD camera \\
\hline
\\
Bacci        & Ref. 0.60-m f/4   & Apogee Alta 1024  \\
Baj          & RC 0.25-m f/8     & SBIG-ST10         \\
Carbognani   & RC 0.81-m f/7.9   & FLI 1001E         \\
Gary         & SC 0.35-m f/10    & SBIG-ST10XME      \\
Oey          & CDK 0.61-m f/6.8  & Apogee U42        \\
\\
\hline
\end{tabular}
\end{center}
\label{ast}
\end{table*}

\begin{table*}
\caption{Observational circumstances for (436724) 2011 UW158. Legend: observer, observation date, the phase angle ($\alpha$), the Phase Angle Bisector (PAB) ecliptic coordinates ($L_{PAB}$, $B_{PAB}$) and lightcurve amplitude A. Note the large lightcurve amplitudes ($\geq 2$ mag) found on some dates. The data for $\alpha$, $L_{PAB}$ and $B_{PAB}$ are from JPL HORIZONS.}
\begin{center}
\begin{tabular}{llllll}
\hline
\\
Obs. & yyyy/mm/dd & Phase (deg) & $L_{PAB}$ (deg) & $B_{PAB}$ (deg) & A (mag) \\
\hline
\\
Gary  & 2015/06/17 & 62.3	& 230.7	& -12.9	& 0.50\\
Gary  & 2015/06/20 & 65.7	& 231.9	& -12.5	& 0.52\\
Oey	  & 2015/07/01 & 79.0	& 236.4	& -9.3	& 0.67\\
Oey	  & 2015/07/02 & 80.3	& 236.8	& -9.3	& 0.67\\
Oey	  & 2015/07/03 & 81.6	& 237.1	& -8.8	& 0.73\\
Oey	  & 2015/07/06 & 85.8	& 238.1	& -6.7	& 0.76\\
Oey	  & 2015/07/08 & 88.9	& 238.6	& -4.7	& 0.71\\
Gary  & 2015/07/08 & 88.9	& 238.6	& -4.7	& 0.70\\
Gary  & 2015/07/12 & 96.0	& 239.4	& 2.3	& 0.70\\
Gary  & 2015/07/20 & 109.3	& 269.9	& 50.6	& 0.92\\
Baj	  & 2015/07/23 & 101.9	& 314.6	& 50.5	& 1.92\\
Bacci &	2015/08/01 & 82.0	& 341.4	& 30.0	& 2.38\\
Carbo. & 2015/08/02 & 80.4	& 342.4	& 29.0	& 1.96\\
Gary  &	2015/08/03 & 78.9	& 343.3	& 28.1	& 1.95\\
Gary  &	2015/08/04 & 77.5	& 344.1	& 27.3	& 2.05\\
Carb. &	2015/08/11 & 68.3	& 348.9	& 23.4	& 1.96\\
Gary  &	2015/08/13 & 65.8	& 350.1	& 22.6	& 1.84\\
Bacci &	2015/08/14 & 64.6	& 350.7	& 22.3	& 1.82\\
Carbo. & 2015/08/19 & 58.6	& 353.4	& 20.8	& 1.62\\
Bacci &	2015/08/20 & 57.4	& 353.9	& 20.5	& 1.65\\
Gary  &	2015/08/29 & 47.0	& 357.9	& 18.6	& 1.46\\
Baj	  & 2015/09/05 & 39.1	& 0.5	& 17.3	& 1.27\\
Gary  &	2015/09/06 & 38.1	& 0.9	& 17.2	& 1.25\\
Baj	  & 2015/09/06 & 38.1	& 0.9	& 17.2	& 1.17\\
Baj	  & 2015/09/08 & 35.9	& 1.5	& 16.8	& 1.34\\
Gary  &	2015/09/15 & 28.3	& 3.7	& 15.7	& 1.07\\
Gary  &	2015/09/16 & 27.8	& 4.0	& 15.5	& 1.08\\
Oey	  & 2015/09/21 & 23.5	& 5.4	& 14.7	& 0.95\\
Gary  &	2015/09/24 & 21.2	& 6.2	& 14.3	& 0.94\\
Gary  &	2015/09/25 & 20.5	& 6.5	& 14.1	& 0.89\\
Gary  &	2015/09/26 & 19.9	& 6.7	& 13.9	& 0.85\\
\\
\hline
\end{tabular}
\end{center}
\label{amp}
\end{table*}

\section{The Cohesionless Spin-Barrier}
\label{sec:1}
\begin{figure}
\centering
\resizebox{0.70\textwidth}{!}{
\includegraphics{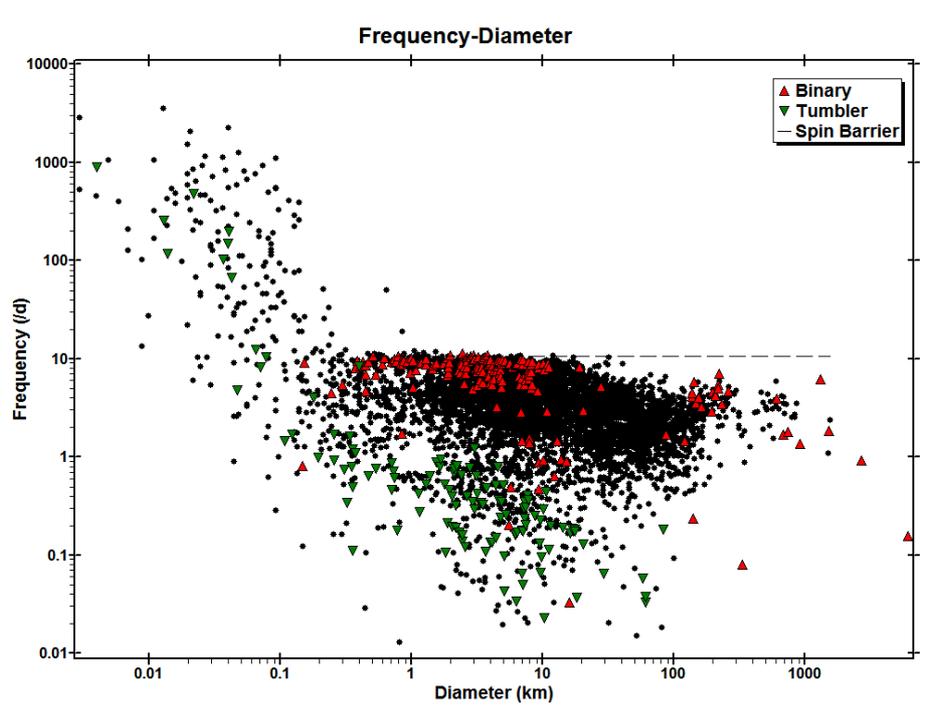}}
\caption{The periods of minor planets plotted as frequency (rotations/day) vs. diameter (km). This figure was drawn from the Asteroid Lightcurve Photometry Database (alcdef.org), held by Brian D. Warner. The spin-barrier threshold is evident for asteroids with diameter between about 0.15 - 10 km. The NEA 2011 UW158 is shown as a large black dot within a red rectangle, representing the range of effective optical diameters corresponding to the smallest to the largest dimension. The radar effective diameter is slightly larger.}
\label{fig:1}       
\end{figure}

Asteroids of size $D \geq 0.15$ km generally do not have rotation periods $P \leq 2.2$ h, a limit known as the cohesionless spin-barrier (Figure \ref{fig:1}). This barrier can be explained by means of the cohesionless rubble-pile structure model \cite{pravec}. According to this model, the asteroids are made up of collisional breakup fragments bound together only by mutual gravitational force. The exceptions to this ``rule'' are very few: (455213) 2001 OE84 \cite{pravec02}, (335433) 2005 UW163 \cite{chan}, and 2011 XA3 \cite{urakawa} are the best known examples. The presence of these ``exotic'' objects, called Large Super-Fast Rotators (LSFRs), see Table \ref{ast_sfrs} for a more complete list, was theorized for the first time by Holsapple \cite{holsapple}. These results have been confirmed and enriched by subsequent theoretical studies, such as that by S\`anchez and Scheeres \cite{sanchez}, in which a model for the origin of the cohesive strength from the regolith has been proposed.\\

\begin{table*}
\caption{The known or candidate LSFRs with name, type, rotation period, discovery year of the SFR state, absolute visual magnitude and references.}
\begin{center}
\begin{tabular}{lclccl}
\hline
\\
Name & Asteroid type & Period (h) & Discovery year & $H_V$ (mag) & References \\
\hline
\\
(455213) 2001 OE84      & Amor              & 0.4865 & 2002 & 17.9 & \cite{pravec02}     \\
2001 FE90               & Apollo-PHA        & 0.4777 & 2009 & 20.7 & \cite{hicks}        \\
2001 VF2                & Amor              & 1.39   & 2011 & 20.3 & \cite{hergenrother} \\
(335433) 2005 UW163     & MBA               & 1.290  & 2014 & 17.7 & \cite{chan}         \\
2011 XA3                & Apollo-PHA        & 0.730  & 2014 & 20.5 & \cite{urakawa}      \\
2014 VQ                 & Amor              & 0.1160 & 2014 & 20.2 & \cite{carbognani_c} \\
\\
\hline
\end{tabular}
\end{center}
\label{ast_sfrs}
\end{table*}

These last authors have explored the hypothesis that, thanks to the small Van der Waals forces between the interstitial regolith grains, an asteroid with rubble-pile structure has a non-zero cohesive strength that enables it to survive even below the spin-barrier value. In practice, the interstitial grains would act as a kind of ``glue'' between the larger block sizes. The final result of the model is that the limit value of the spin rate depends on the size of the asteroid according to the following relation:

\begin{equation}
\omega^{2} \leq {\omega_0}^{2}+\frac{\sigma_{Y}}{a^2\rho}
\label{cohesion}
\end{equation}

\noindent In Eq. (\ref{cohesion}) $\omega = 1/P$ (where $P$ is the rotation period), $\rho$ is the mean density, $a$ is the semi-major axis of the ellipsoid representing the asteroid, $\omega_0$ is the cohesionless spin-barrier value and $\sigma_Y$ is the cohesive strength expressed in force for unit area i.e. a pressure. A similar result was obtained by Holsapple \cite{holsapple}. Equation (\ref{cohesion}) shows that for a given cohesive strength the spin rate to disrupt a rubble-pile asteroid increases for smaller or less dense bodies \cite{sanchez}. Even considering the high value known for $\sigma_Y$, as the 3 kPa of the Moon regolith, the presence of cohesive strength begins to matter only for objects with a diameter below 10 km. So, for small bodies (about $D \leq 10$ km) with rubble-pile structure, the presence of even a very small amount of cohesive strength allows more rapid spin than the simple cohesionless spin-barrier model.

\section{The Discovery of the Fast Rotation Period}
\label{sec:2}
\begin{figure}
\centering
\resizebox{0.70\textwidth}{!}{
\includegraphics{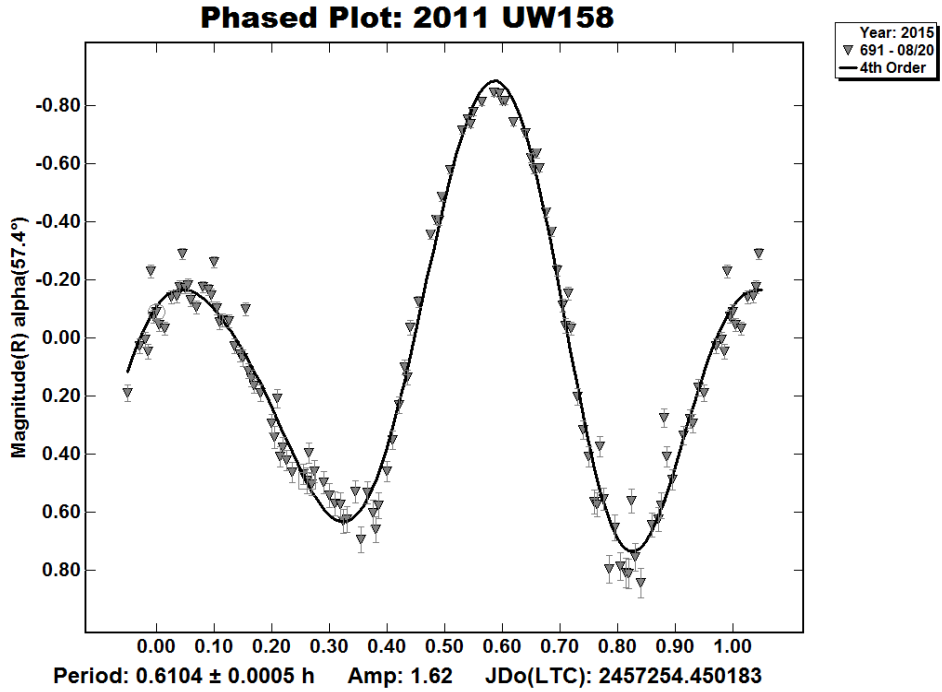}}
\caption{The lightcurve of 2011 UW158 on 2015 Aug 20 (A. Carbognani).}
\label{fig:2}       
\end{figure}

The asteroid 2011 UW158 was first observed by Gary with unfiltered CCD images calibrated using $r'$-mag's of APASS stars in the UCAC4
catalog \cite{gary}. APASS means ``AAVSO Photometric All-Sky Survey'', this catalog contains photometry for 60 million objects in about 99\% of the sky. The 5-band photometry of the APASS is based on Johnson B, V and Sloan $g'$ (SG), $r'$ (SR), $i'$ (SI) filters, and in B, V and SR is a valid reference for asteroids photometry from about 10th to about 15th mag \cite{carbognani_b}. Thanks to these observations a synodic rotation period of only 36.66 minutes was first found by Gary on 2015 June 17 and independently by Oey on July 1. On July 18 the radar observations have given a value of $ 36 \pm 3$ min, in good agreement with the optical result \cite{ipatov}. The lightcurve for this asteroid appear with a typically bimodal trend, with two maxima and two minima, see Figure \ref{fig:2} for a representative example. With further observations, and using the phase curve model of Belskaya and Schevchenko \cite{schev}, also the geometric albedo $p_V$ and size was estimated. The results are $p_V\approx 0.39\pm 0.09$ and $D\approx 220\pm 40 $ m \cite{gary}. The high albedo value suggest an S-type or E-type asteroid \cite{belskaya}. The visible extents of the asteroid in the radar images suggest an elongated object with dimensions of about $600\times 300$ m \cite{naidu}, so with an effective diameter of about 380 m, even larger than optical observations. With this diameter value coupled with the fast rotation period, 2011 UW158 is a very good LSFR candidate (Figure \ref{fig:1}).

\section{The Amplitude-Phase Relationship}
\label{sec:3b}
From the photometric observations we had not only the rotation period but also the lightcurve amplitude (see Table \ref{amp}). While the rotation period is a fixed parameter, the amplitude depends both on the object's shape and on the aspect angle at the time of observations. However, the amplitude can be affected also by the phase angle $\alpha$ and there is an APR (Amplitude-Phase Relationship) \cite{zappala}. The APR can be fit by a linear equation of the form:

\begin{equation}
A\left(\alpha\right) = A\left(0^{\circ}\right)\left(1+ m\alpha\right)
\label{APR}
\end{equation}

\begin{figure}
\centering
\resizebox{0.60\textwidth}{!}{
\includegraphics{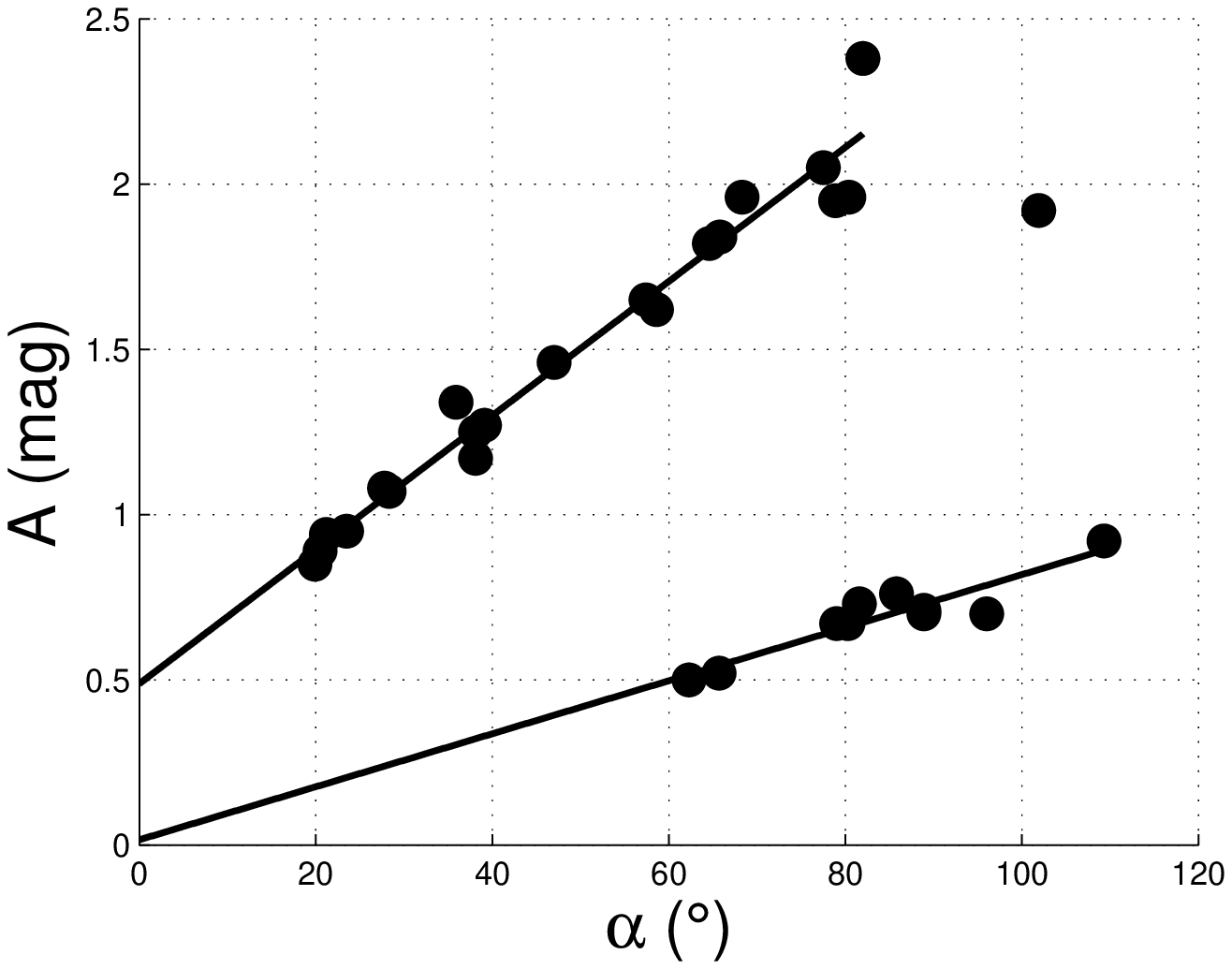}}
\caption{The amplitude-phase relationship for 2011 UW158. The lower plot refers to the pre-flyby, the upper one to the post-flyby with Earth.}
\label{fig:3a}       
\end{figure}

\noindent In eq. (\ref{APR}), $A\left(\alpha\right)$ is the amplitude (in mag) at the phase angle $\alpha$ (in deg), while $m$ is a constant with dimension 1/deg that depends on the taxonomic type, i.e. $m = 0.030$ for S-type, $m = 0.015$ for C-type and $m = 0.013$ for M-type \cite{zappala}. In our case the APR relation is shown in Figure \ref{fig:3a}. Note that there are two different plots. The lower plot concerns the pre-flyby with Earth, the upper one the post-flyby period. A linear fit with the least square method of eq. \ref{APR} gives, for the upper plot: $A\left(0^{\circ}\right) = 0.49 \pm 0.05$ mag, $A\left(0^{\circ}\right) m = 0.0203 \pm 0.0009$ mag/deg and $m = 0.041 \pm 0.005\, \hbox{deg}^{-1}$. This last value is compatible with an S-type asteroid and is thus in discrete agreement with the geometric albedo found by Gary \cite{gary}. A linear fit of the lower plot gives: $A\left(0^{\circ}\right) = 0.016 \pm 0.096$ mag, $A\left(0^{\circ}\right) m = 0.008 \pm 0.001$ mag/deg and $m = 0.5 \pm 3\, \hbox{deg}^{-1}$. In this last case the $m$ uncertainty is too large to be useful, i.e. $m$ cannot be determined from the present set of measurements. These different results between the two plots are due to the fact that the pre-flyby plot has the points in a minor phase-angle range than the post-flyby. So the corresponding $A\left(0^{\circ}\right)$ value has a much greater uncertainty and, with errors propagation, we have a very high uncertainty on $m$.

\section{Pole Search}
\label{sec:4}
\begin{figure}
\centering
\resizebox{0.70\textwidth}{!}{
\includegraphics{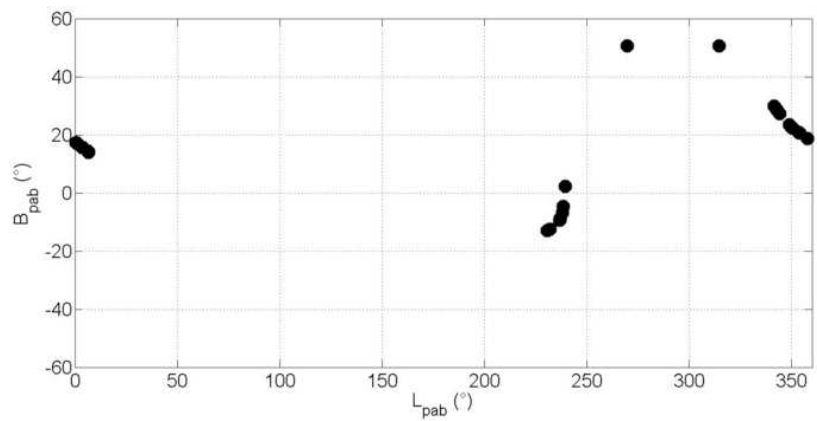}}
\caption{Distribution of the phase angle bisector (PAB) for 2011 UW158.}
\label{fig:4}       
\end{figure}

Our main purpose with 2011 UW158 was to determine the pole of rotation and convex shape using the standard lightcurve (LC) inversion method \cite{kaasalainen1} \cite{kaasalainen2}. In most cases, it is not possible to get a reasonable solution for a pole using LC inversion with photometric observations from one apparition. In our case the range of observed phase angle is $62^{\circ}$ to $109^{\circ}$ and $109^{\circ}$ to $20^{\circ}$ while the amplitudes of the angle bisector are $\Delta$LPAB = $137^{\circ}$ and $\Delta$BPAB = $64^{\circ}$, sufficiently broad for trying to determine pole orientation and shape (Figure \ref{fig:4}). The LC inversion process was performed using MPO LCInvert v11.1.0.2 (Bdw Publishing), which implements the core algorithms developed by Kaasalainen and then converted to C language by Josef Durech.

\begin{figure}
\centering
\resizebox{0.70\textwidth}{!}{
\includegraphics{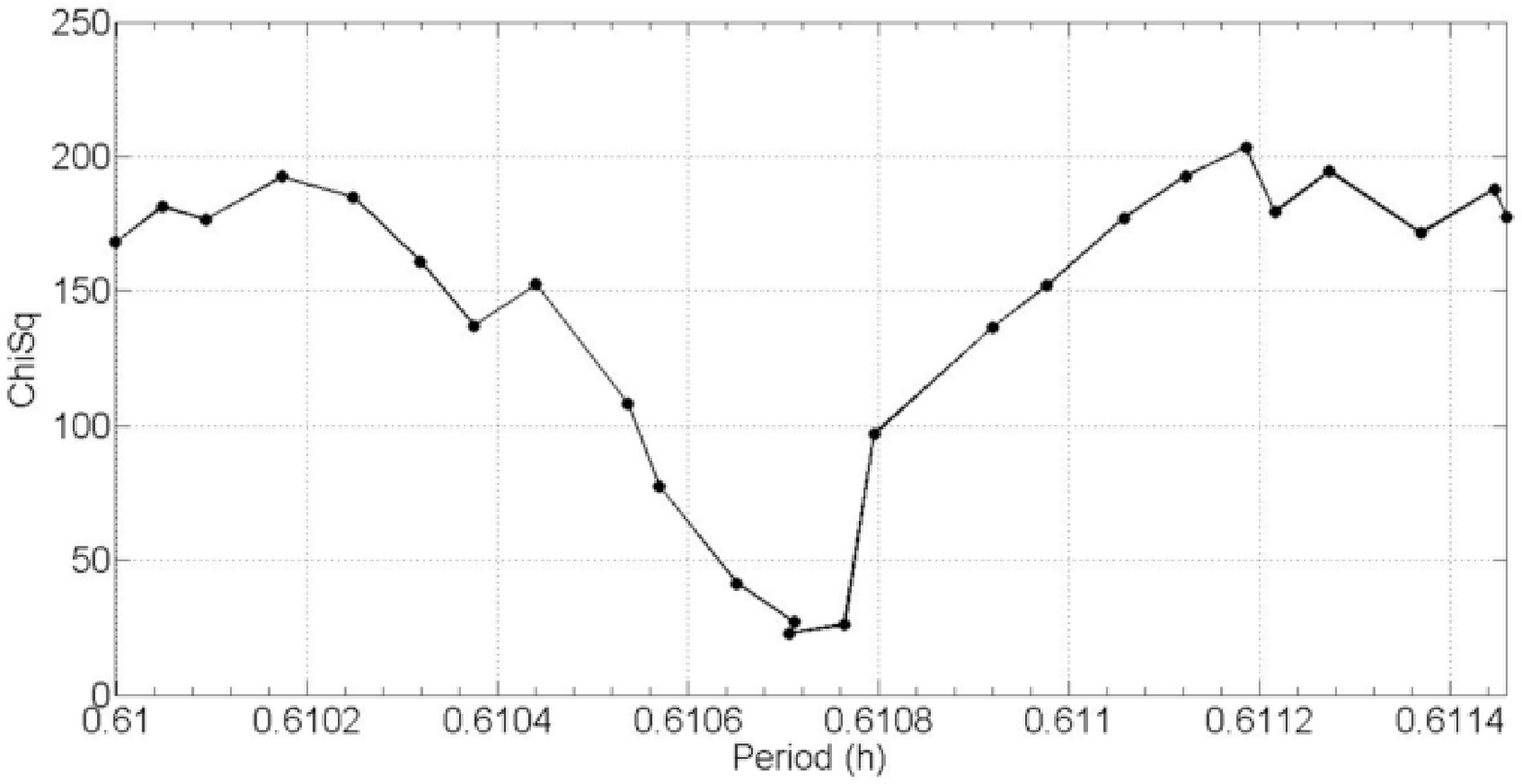}}
\caption{Sum of ChiSq ($\chi^2$) vs period for 2011 UW158 \cite{carbognani}.}
\label{fig:5}       
\end{figure}

The inversion process started by finding the sidereal rotation period of the asteroid \cite{carbognani}. A search in MPO LCInvert was confined to the range from 0.6100 to 0.6115 h, which includes the synodic period found in the single phased LC, with weight 0.5. However, inclusion of all observations in table \ref{amp} leads to $\chi^2$ values that are quite high. After some tests, we found that by restricting observations to those by Gary (in this way the range of the phase angle remains unchanged) and those before Aug 15 for the other observers, the $\chi^2$ values were reduced to reasonable values. The search process found an isolated, deep, and flat minimum in the
plot of $\chi^2$ vs. sidereal period (Figure \ref{fig:5}). A renormalization was not necessary since reduced $\chi^2 \sim 1.0$ (i.e., N = 24 and sum $\chi^2$ is also ~24). The minimum appears to be asymmetrical, i.e. the descending branch is less steep than the ascending branch. For this reason we assumed the value of the point to the right, 0.6107643 h, for the starting period in the pole search.

\begin{figure}
\centering
\resizebox{0.70\textwidth}{!}{
\includegraphics{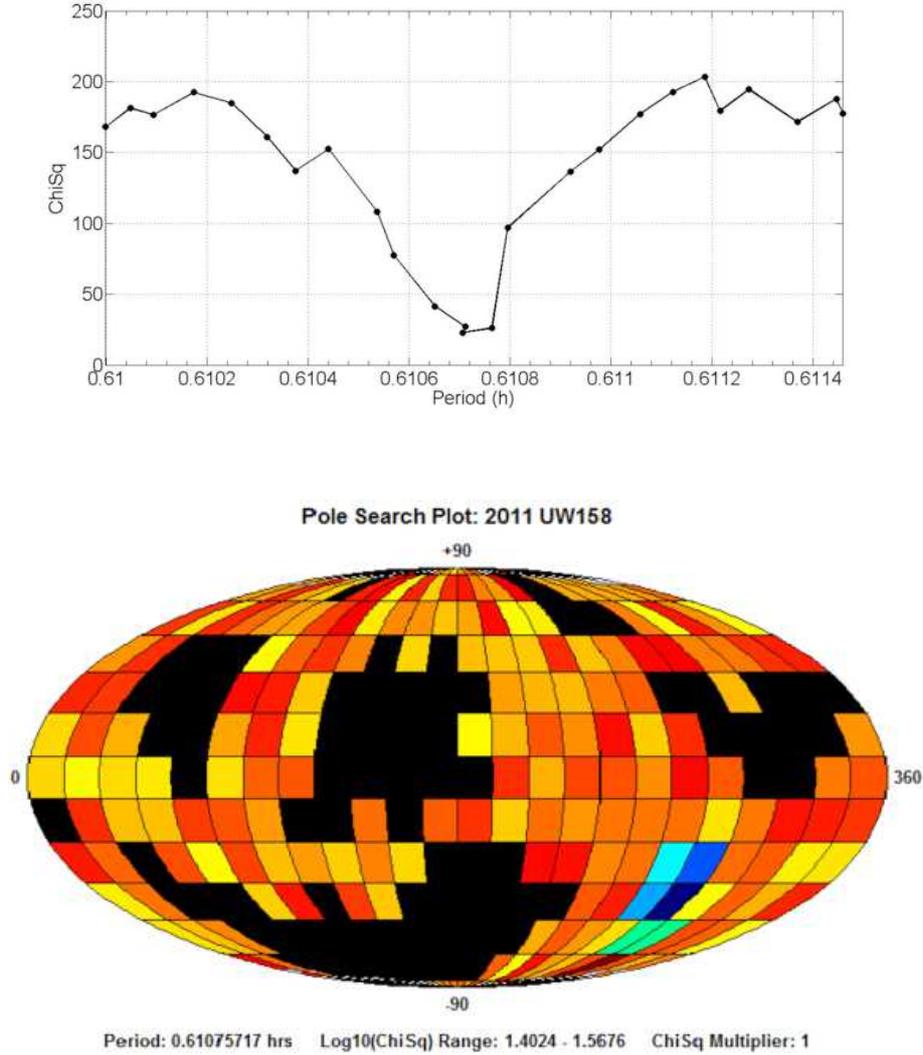}}
\caption{Results of the "medium" pole search as a map of $\chi^2$ on the ecliptic sky. The deep blue region represents the pole location with the lowest sum of Chi-square which increases as the color goes from light blue to green to yellow to orange and finally to deep red. Black regions indicate where the code produced an invalid result i.e. NaN, not a number \cite{carbognani}.}
\label{fig:6}       
\end{figure}

For the pole orientation search, we started using the ``Medium'' search option in LCInvert (312 fixed pole positions with $15^{\circ}$ longitude-latitude steps). The previously found sidereal period was set to ``float'' and the weight parameter = 0.8. The pole search found one cluster of solutions centered around ecliptic coordinates $\lambda = 285^{\circ}$ and $\beta = –45^{\circ}$ with a sidereal period $P = 0.61075717$ h. Figure \ref{fig:6} shows the distribution of $log(\chi^2)$ values. A final search for a spin axis solution was made using the lowest value in this island. Here the longitude and latitude are allowed to float, as was the period. The spin axis parameters were then used to generate a final shape and spin axis model. Refining the pole search, using the ``Fin'' option of LCInvert software (49 fixed pole steps with $10^{\circ}$ longitude-latitude pairs) and the previous period/longitude/latitude set to ``float'', we found the best solution to be ecliptic coordinates $\lambda = 290^{\circ} \pm 3^{\circ}$ and $\beta = –39^{\circ} \pm 2^{\circ}$ (near the star alpha Pavonis), with an averaged sidereal period $P_s = 0.610752 \pm 0.000001$ h. The uncertainty in $\lambda$, $\beta$, and sidereal period are chosen to be the mean standard deviation of the 49 single solution of the fine pole search. Since the ecliptic latitude of the rotations axis is negative, the asteroid has a retrograde rotation, i.e., it rotates clockwise when viewed from the ecliptic north pole. This result seems to be confirmed by radar observations, see Figure\ref{fig:9} \cite{ipatov}. The study of the prograde-retrograde spin distribution of near-Earth asteroids is important, e.g., for the model of orbital drift of these bodies \cite{spina}. Of course this is a preliminary solution, but our confidence in the final solution is bolstered by the fact that the first half of the data gave only two possible solutions, one of which is the same solution using all data \cite{carbognani}.

\begin{figure}
\centering
\resizebox{0.70\textwidth}{!}{
\includegraphics{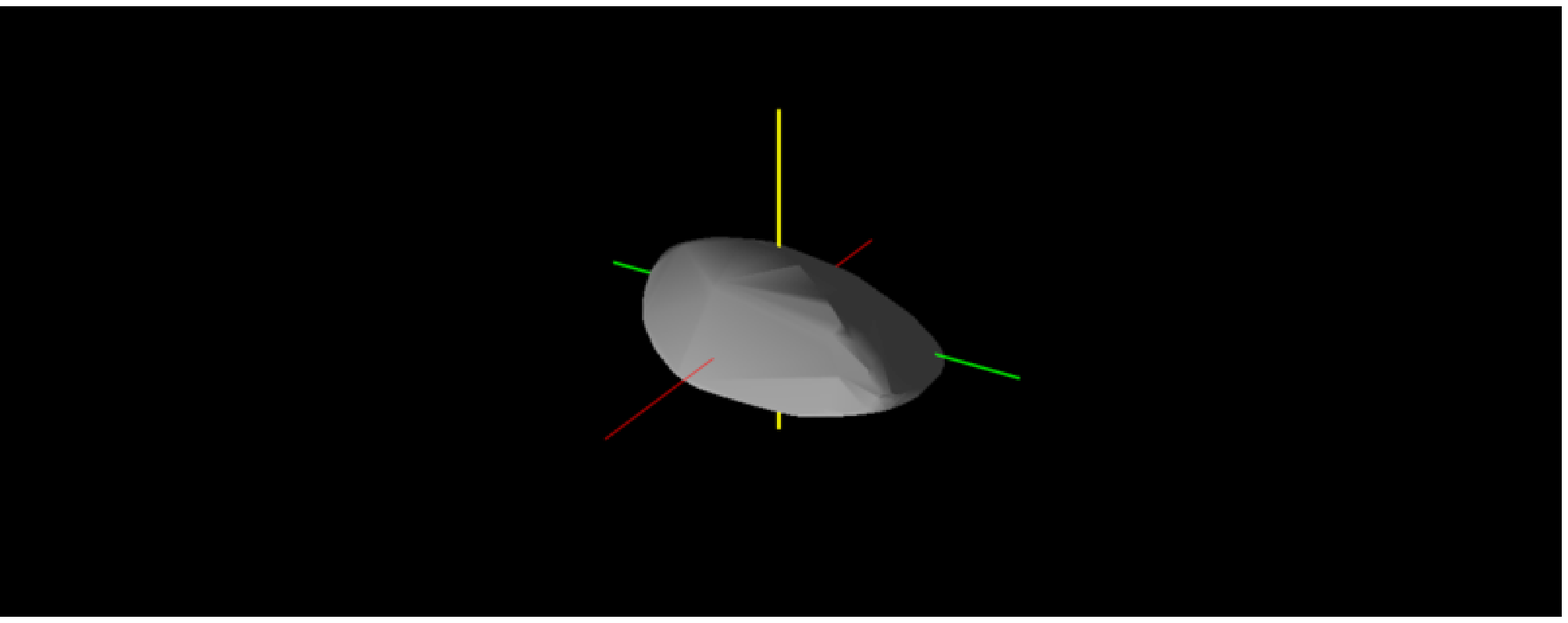}}
\caption{The 3D best model for 2011 UW158, with pole in $\lambda = +295°$, $\beta = –40°$ (a/b=1.3, a/c=2.3, b/c=1.7).}
\label{fig:7}       
\end{figure}

\begin{figure}
\centering
\resizebox{0.70\textwidth}{!}{
\includegraphics{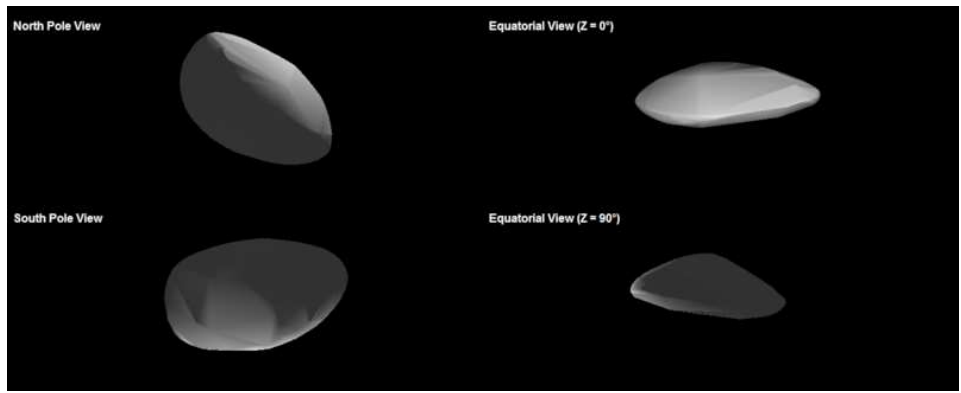}}
\caption{Multiview of the best 3D model for 2011 UW158.}
\label{fig:7a}       
\end{figure}


\section{Shape Model}
\label{sec:5}
\begin{figure}
\centering
\resizebox{0.70\textwidth}{!}{
\includegraphics{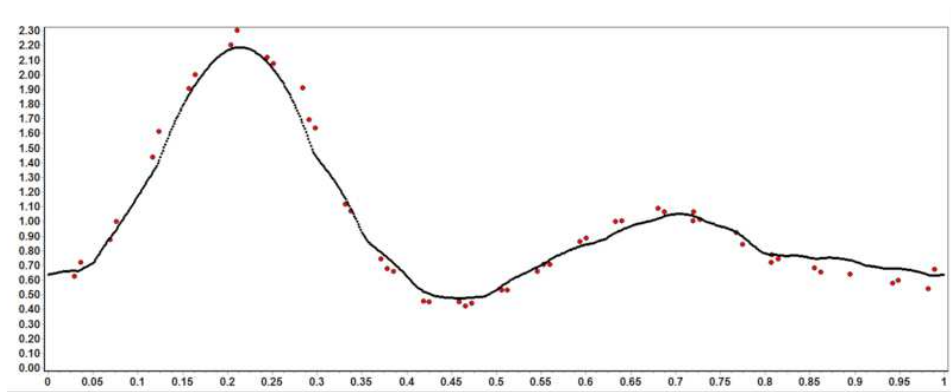}}
\caption{Comparison between the shape model synthetic LC with the observed LC on Aug 14 2015.}
\label{fig:6a}       
\end{figure}

The best shape model for this asteroid (n. 24 in our data processing), shows a rather elongated object in rotation around the minor axis (Figure \ref{fig:7} and Figure \ref{fig:7a}). This result is consistent from the physical point of view, and in agreement with the large LC amplitudes ($>2$ mag) found on some observing dates. This 3D shape is also in good agreement with the radar observations \cite{naidu}, \cite{ipatov}. We tested the shape model by comparing synthetic lightcurves with observed ones (Figure \ref{fig:6a}). The shape model produces synthetic lightcurves that are in very good agreement with the observed lightcurves \cite{carbognani}.


\section{Conclusions}
\label{sec:6}
We have carried out a brief review of the cohesionless rubble-pile model and seen a list of objects that violate the cohesionless spin-barrier. This discrepancy is probably due to the cohesive strength exerted by the regolith on the blocks that compose the asteroid. In search of other rare asteroids of this type, we have presented the optical results obtained in 2015 about the Apollo-PHA (436724) 2011 UW158, a very good candidate to be a LSFR. We have seen the fast rotation period discovery circumstances, the lightcurve, the amplitude-phase relationship, the spin axis determination and a 3D model in good agreement with radar observations. An explicit search for spin changes, before and after the Earth flyby, was not made because pre-flyby data are not sufficient to determine the spin.

\section{Acknowledgements}
\label{sec:7}
This research has made use of the NASA's Astrophysics Data System, JPL Small-Body Database Browser and JPL's HORIZONS system. We also use APASS database, located at the AAVSO web site. Funding for APASS has been provided by the Robert Martin Ayers Sciences Fund. This research has also made use of the VizieR catalogue access tool, CDS, Strasbourg, France. The original description of the VizieR service was published in Astron. Astrophys. Suppl. Ser. \textbf{143}, 23 (2000). The Astronomical Observatory of the Autonomous Region of the Aosta Valley (OAVdA) is managed by the Fondazione Clément Fillietroz-ONLUS, which is supported by the Regional Government of the Aosta Valley, the Town Municipality of Nus and the "Unité des Communes valdôtaines Mont-Émilius". The research was partially funded by a 2016 "Research and Education" grant from Fondazione CRT. Research at OAVdA was also supported by the 2013 Shoemaker NEO Grant. Work at the Blue Mountains Observatory was supported by the 2015 Shoemaker NEO Grant.

%
%
%

%
%

\end{document}